\documentclass[conference]{IEEEtran}
\IEEEoverridecommandlockouts
\usepackage{cite}
\usepackage{amsmath,amssymb,amsfonts}
\usepackage{algorithmic}
\usepackage{graphicx}
\usepackage{textcomp}
\usepackage{xcolor}
\usepackage{subfigure}
\usepackage{enumitem}
\usepackage[linesnumbered,ruled,vlined]{algorithm2e}
\usepackage{booktabs}
\usepackage{pifont}
\usepackage[hidelinks]{hyperref}
\usepackage{caption}
\renewcommand{\thetable}{\arabic{table}}

\graphicspath{{fig/}}

\newcommand{\ours}{BASTS\xspace}
\newcommand{\oursnormal}{BASTS$_{b128}$\xspace}
\newcommand{\ourssmallbs}{BASTS$_{b64}$\xspace}
\newcommand{\xbf}{\mathbf{x}}
\newcommand{\ybf}{\mathbf{y}}
\newcommand{\ebf}{\mathbf{e}}
\newcommand{\Ncal}{\mathcal{N}} 
\newcommand{\Ocal}{\mathcal{O}} 
\newcommand{\Wcal}{\mathcal{W}} 
\newcommand{\Ucal}{\mathcal{U}} 
\newcommand{\Tcal}{\mathcal{T}} 
\newcommand{\Vcal}{\mathcal{V}} 
\newcommand{\Ecal}{\mathcal{E}} 
\newcommand{\Rcal}{\mathcal{R}} 
\newcommand{\Ccal}{\mathcal{C}} 
\newcommand{\ibf}{\mathbf{i}}
\newcommand{\fbf}{\mathbf{f}}
\newcommand{\obf}{\mathbf{o}}
\newcommand{\hbf}{\mathbf{h}}
\newcommand{\mbf}{\mathbf{m}} 
\newcommand{\ubf}{\mathbf{u}}
\newcommand{\bbf}{\mathbf{b}}
\newcommand{\Wbf}{\mathbf{W}}
\newcommand{\Ubf}{\mathbf{U}}

\newcommand{\emaildot}{\makebox[0.2em]{\scalebox{.25}{\textbullet}}}

\begin{document}

\bstctlcite{IEEEexample:BSTcontrol}

\title{Improving Code Summarization with Block-wise Abstract Syntax Tree Splitting}

\author{
	\IEEEauthorblockN{Chen Lin, Zhichao Ouyang, Junqing Zhuang, Jianqiang Chen, Hui Li$^{*}$\thanks{* Hui Li is the corresponding author.}, Rongxin Wu}
	\IEEEauthorblockA{
		\textit{Xiamen University} \\
		Xiamen, China \\
		\{chenlin, hui, wurongxin\}@xmu\emaildot edu\emaildot cn, \{ouyangzhichao, zhuangjq, jqchen\}@stu\emaildot xmu\emaildot edu\emaildot cn 
	}
}

\maketitle

\begin{abstract}
Automatic code summarization frees software developers from the heavy burden of manual commenting and benefits software development and maintenance. Abstract Syntax Tree (AST), which depicts the source code's syntactic structure, has been incorporated to guide the generation of code summaries. However, existing AST based methods suffer from the difficulty of training and generate inadequate code summaries. In this paper, we present the Block-wise Abstract Syntax Tree Splitting method (BASTS for short), which fully utilizes the rich tree-form syntax structure in ASTs, for improving code summarization. BASTS splits the code of a method based on the blocks in the dominator tree of the Control Flow Graph, and generates a split AST for each code split. Each split AST is then modeled by a Tree-LSTM using a pre-training strategy to capture local non-linear syntax encoding. The learned syntax encoding is combined with code encoding, and fed into Transformer to generate high-quality code summaries. Comprehensive experiments on benchmarks have demonstrated that BASTS significantly outperforms state-of-the-art approaches in terms of various evaluation metrics. To facilitate reproducibility, our implementation is available at \url{https://github.com/XMUDM/BASTS}.
\end{abstract}


\section{Introduction}\label{sec:intro}

Human-readable code summaries (i.e., code comments) are the key to enhance the comprehension of source code in software development and maintenance~\cite{AllamanisBDS18,abs-2002-05442}.
As software projects grow rapidly in scale and complexity, manual commenting becomes costly for developers.  
There has been increasing interest in \emph{automatic code summarization}~\cite{abs-1909-04352}, which automatically generates short and accurate textual descriptions for software entities and frees software developers from the tremendous workload of writing comprehensible comments.

Early works on code summarization select a subset of the statements and
keywords from the code to generate comments~\cite{abs-1909-04352}. 
However, the grammar of programming languages strictly follows specific syntaxes, which is not easy to model using code tokens only. It is critical to encode both the meaning
of each code token and the syntax structure when generating code summarization. 
Abstract
Syntax Tree (AST), which depicts the hierarchical abstract syntactic structure of source code, is therefore commonly
harnessed together with code tokens to improve the quality of code summarizations~\cite{Hu2018Deep,Alon2019code2seq,Kovalenko2019PathMiner}. 
Nevertheless, it is infeasible to directly adopt the methods studying the syntax tree in Natural Language Processing (NLP) and apply tree-aware neural networks (e.g., Tree-LSTM~\cite{Sheng2015Improved}) in the code summarization task. The reason is that ASTs are quite different compared to NLP syntax trees.
The size of an AST in terms of tree depth is very large, leading to a heavy encoder which is difficult and slow to train. For instance, the maximum AST depth of the two datasets in our experiments (Sec.~\ref{sec:experiment}) is 76 (Java) and 112 (Python), which is rare for NLP syntax trees and weakens the capability of neural networks to capture the complex semantics~\cite{ZhuSG15}. Graph Neural Network (GNN)~\cite{abs-1901-00596} based methods have the similar problem of the heavy encoder and their results do not show significant improvements over other methods in code summarization~\cite{LeClairHWM20}.

\begin{figure*}[htbp]
\begin{center}
\includegraphics[width=0.88\textwidth]{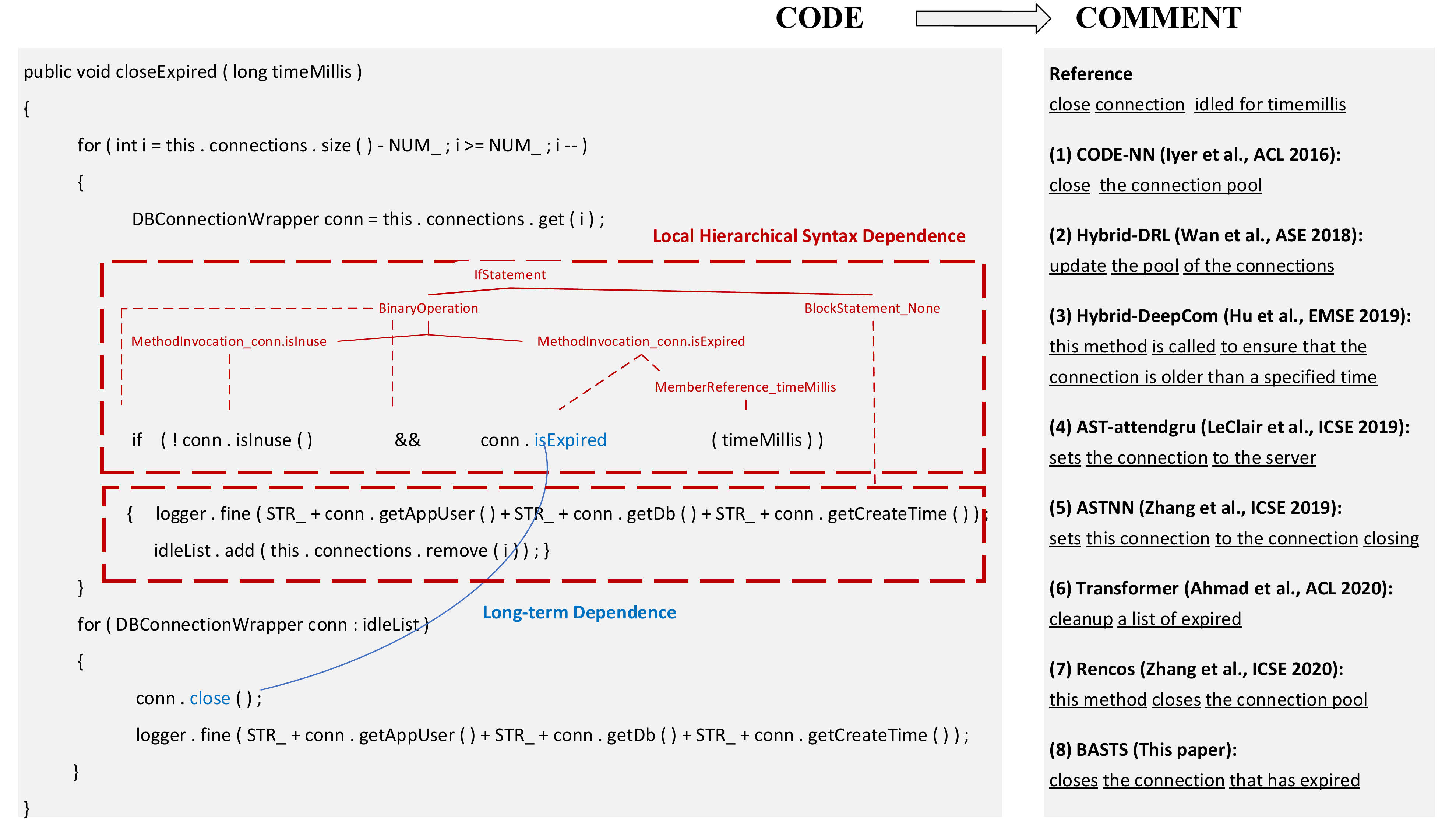}
\caption{Outputs of different automatic code summarization methods for test sample $\#1445$ in the Java dataset }
\label{fig:runningexample}
\end{center}
\vspace{-18pt}
\end{figure*}

To overcome the challenge of training over the entire AST, a great number of 
methods are proposed. But they all produce inadequate
summaries. We now use the example from the Java dataset in our experiments to
illustrate their limitations. In Fig.~\ref{fig:runningexample}, reference
(i.e., gold standard summary) fully reveals the hierarchical syntax in the AST
and the long-term dependence in the source code. For example, in the
hierarchical structure of the AST, ``\emph{timemillis}" is linked to
``\emph{expire}'' to form a sub-clause, where the sub-clause is in turn linked
to a connection status in an ``\emph{if}'' clause. The ``\emph{if}'' clause
triggers the ``\emph{close}'' action via the long-term dependence. 
Accordingly, in the reference, ``\emph{timemillis}'' in the ``\emph{idle}''
clause complements ``\emph{connection}''. It also complements ``\emph{close}''
with ``\emph{connection (if it has) idled for timemillis}''. Next, let us see the results from contemporary code
summarization methods which can be roughly divided into three categories:
\begin{enumerate}[leftmargin=13pt] 
\item Most works serialize ASTs and then adopt a sequential neural
model. Specifically, an AST can be serialized as a single sequence of
nodes~\cite{Hu2018Deep,LeClairJM19,Hu2019HDeep,ZhangW00020} via traversal, or a collection of paths~\cite{Alon2019code2seq,Kovalenko2019PathMiner}
(i.e., several sequences of nodes). Serialized ASTs can be
easily modeled, as modeling sequential natural language
tokens~\cite{YoungHPC18}, by sequential neural models such as Recurrent
Neural Network (RNN)~\cite{rumelhart1986learning}, Long Short Term Memory
(LSTM)~\cite{HochreiterS97}, Gated Recurrent Unit
(GRU)~\cite{ChoMGBBSB14} or Transformer~\cite{Vaswani2017Attention}. However, such a strategy encodes only the linear
order syntax structure and inevitably loses \textit{hierarchical syntax
information}. 
In Fig.~\ref{fig:runningexample},
AST-attendgru~\cite{LeClairJM19}, Hybrid-DeepCom~\cite{Hu2019HDeep} and
Rencos~\cite{ZhangW00020} which learn from serialized ASTs, are unable to
capture the relations (including ``\emph{connection is expired}'' or
``\emph{close the connection}'') correctly and perform even worse than
Code-NN~\cite{Iyer2016Summarizing} which does not consider ASTs.
\item A few works simplify the tree structure of ASTs, e.g., transform an AST to a
binary tree~\cite{Wan2018Improving}. However, they change the
original semantics of code and bring information
loss~\cite{ZhangWZ0WL19}. 
In Fig.~\ref{fig:runningexample},
Hybrid-DRL~\cite{Wan2018Improving}, which learns from binarized AST, is unable
to capture the long-term dependence between ``\textit{connection is expired}"
and ``\textit{close the connection}" as well as the relation between
``\emph{close}'' and ``\emph{connection}''.
\item Last direction is most related to our work. ASTNN~\cite{ZhangWZ0WL19} splits an AST into small statement trees, and each tree consists of AST nodes of one statement. The statement tree is further encoded by Recursive Neural Network (RvNN)~\cite{SocherLNM11}. This way, the difficulty of training is reduced. Nevertheless, such a strategy forces RvNN to focus on one statement each time, as it only see one statement in each split AST. The long-term dependence, which is beyond the boundary of one statement, cannot be captured by RvNN only. If the subsequent sequential neural model fails to fuse multiple statement trees well, the generated summary will solely contain information from one statement. For instance, in Fig.~\ref{fig:runningexample}, ASTNN fails to capture the dependence between ``\emph{connection is expired}'' and ``\emph{close the connection}''. More analysis can be found in Sec.~\ref{sec:contri}.
\end{enumerate}

In this paper, we propose the \underline{B}lock-wise \underline{A}bstract \underline{S}yntax \underline{T}ree \underline{S}plitting (\ours) to address aforementioned issues.
Specifically, we split the code of a method based on the \emph{blocks} in the dominator tree~\cite{prosser1959applications} of the Control Flow Graph of the method~\cite{allen1970control}, instead of statements in the code~\cite{ZhangW00020}, to capture the information beyond the boundary of one statement.
After splitting, for each split code, we generate a \textit{split AST}, which is modeled by a Tree-LSTM. 
Furthermore, inspired by fine-tuning a pre-trained model on specific tasks in NLP~\cite{DevlinCLT19}, we adopt pre-training to learn split AST representations by predicting next split AST in the dominator tree. 
A high-quality comment requires understanding of both local dependence and long-term dependence. Therefore, 
we combine the split AST representations with the source code representations in the Transformer architecture~\cite{Vaswani2017Attention} which has the ability to model both types of dependences. 
As shown in Fig.~\ref{fig:runningexample}, only \ours accurately recovers all relations in the reference. 

The proposed \ours framework has three advantages:
\begin{enumerate}[leftmargin=13pt]
\item Through code splitting, each AST is reduced to an affordable tree propagation space for building the Tree-LSTM, and each block still retains sufficient information. Thus, \ours is efficient and scalable for large programs. 
\item Pre-training involves only the syntax representations. Hence, by first pre-training and then fine-tuning the syntax representations with token representations, \ours saves computational resources, compared to building both representations from scratch. Moreover, the pre-training is unsupervised, which is endowed with the potential to handle sparse training data without sufficient code comments. 
\item \ours is effective since it learns both long-term dependencies on tokens (i.e., with the attention mechanism of Transformer) and local hierarchical dependencies on syntax structures (i.e., with Tree-LSTM in the local context). 
\end{enumerate}

\begin{figure*}[t]
\begin{center}
\includegraphics[width=0.9\textwidth]{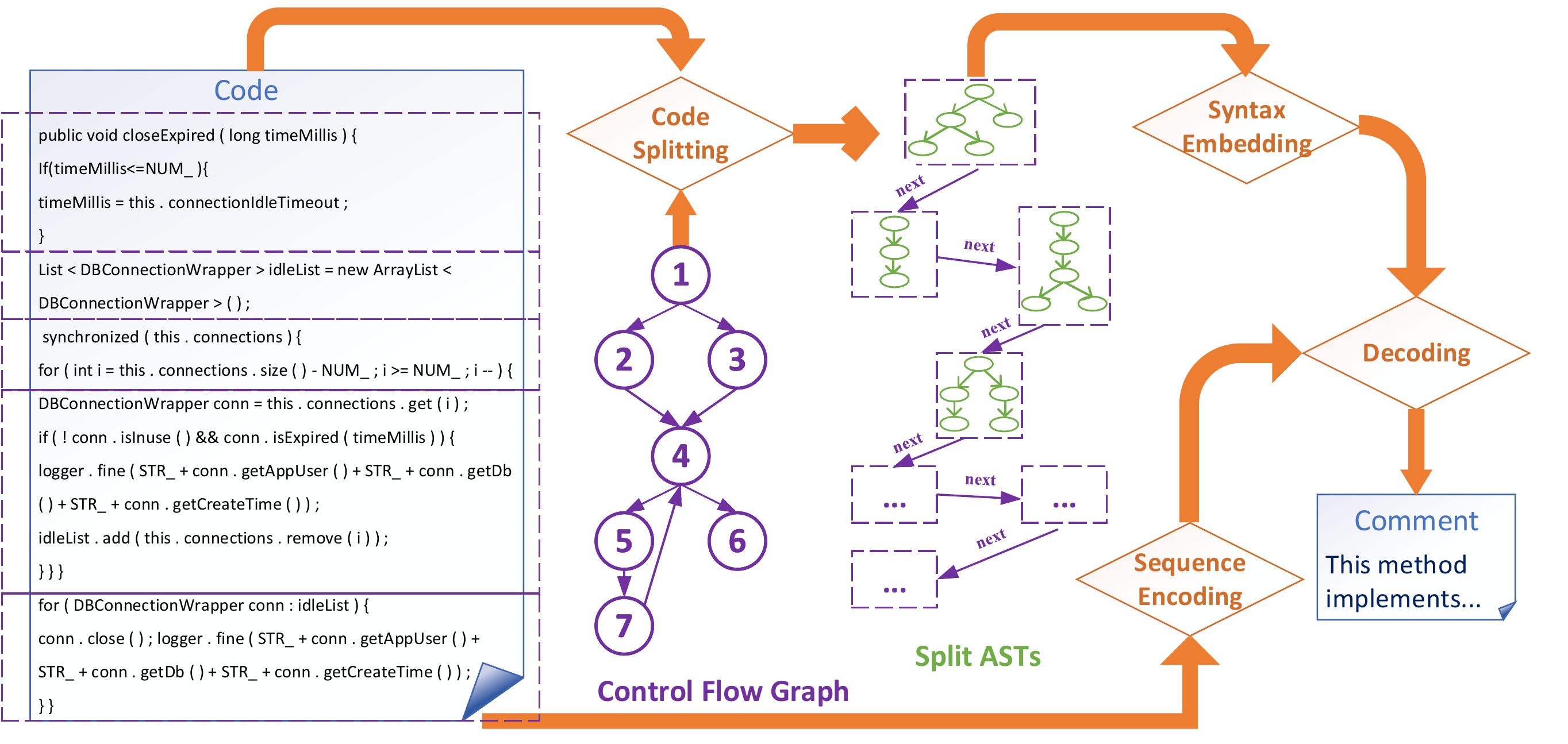}
\caption{Overview of \ours. The CFG and split ASTs are simplified for illustration purpose.}
\label{fig:framework}
\end{center}
\vspace{-10pt}
\end{figure*}

In summary, the contributions of this paper are:
\begin{enumerate}[leftmargin=13pt]
\item We present \ours which incorporates normal tree structure of AST rather than serialized AST or binary AST to avoid information loss in code representation learning.
\item We improve automatic code summarization by using block-wise AST splitting, pre-training and Transformer to effectively learn local and hierarchical syntax dependences.   
\item We conduct comprehensive experiments on benchmarks for code summarization to verify the effectiveness of \ours.  
\end{enumerate}

The rest of the paper is organized as follows.  We present an overview of
\ours in Sec.~\ref{sec:overview}. We illustrate the block-wise code splitting
in Sec.~\ref{sec:code_split}. We describe in detail the neural models of
pre-training syntax representations in Sec.~\ref{sec:pretrain} and code
summarization in Sec.~\ref{sec:summarization}. We report and analyze the
experimental results in Sec.~\ref{sec:experiment}. We elaborate on related
work in Sec.~\ref{sec:rel}. Finally, we conclude our work in
Sec.~\ref{sec:conclusion}.

\section{Overview}
\label{sec:overview}

Given a program (e.g., code fragment for a method or a class), the aim of code summarization is to generate a short paragraph of concise and coherent comments to describe the function of the program. 
Hereafter, unless stated otherwise, we use lower-case letters for variables, upper-case letters for constants, lower-case bold letters for vectors, upper-case bold-face letters for matrices, upper-case calligraphic letters for sets, superscripts for sample index, and subscripts for vector index. 
To be specific, we are given a set of source programs and comments $\Ncal=\{\xbf^n,\ybf^n\}, 1\leq n\leq N$, where $N$ is the number of training instances. 
Each source program is a sequence of tokens, i.e. $\xbf^n=\langle\xbf^n_1,\cdots,\xbf^n_{D_n}\rangle$, where $D_n$ is the number of tokens for the $n$-th sample. 
In the training set, each source program is paired with a comment, which is also a sequence of words, i.e., $\ybf^n=\langle\ybf^n_1,\cdots,\ybf^n_{S_n}\rangle$, where $S_n$ is the number of words for the $n$-th sample. 
The tokens are from the programing language vocabulary and the words are from the natural language vocabulary, i.e., $\xbf^n_d\in \Ocal, 1\leq d\leq D_n$ and $\ybf^n_s \in \Wcal, 1\leq s\leq S_n$. 

\ours consists of three components. 
As shown in Fig.~\ref{fig:framework}, we first pre-process the program by code splitting which is performed according to the Control Flow Graph (CFG), and we generate a set of split ASTs on the split code. 
Then, in \emph{tree encoding}, we pre-train the representations for each split AST to obtain \emph{syntax embedding}, i.e., hierarchical syntax representations of the split AST. 
Finally, the summarization model is trained based on \emph{decoding} input, including the pre-trained syntax embedding in tree encoding, and the \emph{sequence encoding} of source code $\xbf_n$, to output comment $\ybf^n$.

\section{Code Splitting}
\label{sec:code_split}

The goal of code splitting is to facilitate encoding local hierarchical syntax structure. 
We expect that each split code after the splitting would 
provide concise and local context to locate tree-form syntax dependence among code tokens.
In this work, we choose to split the code of a method based on the blocks in the dominator tree~\cite{prosser1959applications} of the CFG of the method~\cite{allen1970control}. 
Following steps show how to generate split code: 
 
\begin{enumerate}[leftmargin=13pt]
\item Construct CFG to capture control flow relationships among statements. In the following, we will omit the sample index and use $G=\big(\Ucal(G), \Ecal(G)\big)$ to denote the CFG for the current training sample. A CFG is a directed graph where each node $u\in\Ucal(G)$ represents a statement, and each edge $e\in\Ecal(G)=(u\rightarrow u')$ represents the control flow from statement $u$ to $u'$. Note that the start node and end node of the CFG are both virtual which do not relate to any real statements. 
\item Construct the dominator tree on the CFG. A dominator tree is a directed graph $DT=\big(\Ucal(DT), \Ecal(DT)\big)$, where $\Ucal(DT)=\Ucal(G)$, and $ \Ecal(DT)\subset  \Ecal(G)$. Each edge is connected, i.e., $e\in\Ecal(DT)=(u\rightarrow u')$ if and only if $u$ dominates $u'$. $u$ dominates $u'$ if and only if every path in CFG from the start node to $u'$ goes through $u$. 
\item Partition the dominator tree into the blocks as split code. 
Each block in the dominator tree is a set of consecutive nodes containing no branches except at the very end.
Thus we remove every edge $e\in\Ecal(DT)=(u\rightarrow u')$ if $u'$ has more than one in-coming edges, or $u$ has more than one out-going edges. In the end, we will have multiple disconnected subgraphs, each corresponds to a set of statements in the code, i.e., a piece of split code. 
\item We add the method declaration statement in the original complete code to the beginning of each split code, and use sophisticated tool to extract AST for each split code, i.e., \textit{split AST}. Details of the AST generation tool will be described in Sec.~\ref{sec:exp_setup}. 
\end{enumerate}

\begin{figure}
  \centering
  \subfigure[Control flow graph]{
    \label{fig:cfg}
    \includegraphics[width=0.8\columnwidth]{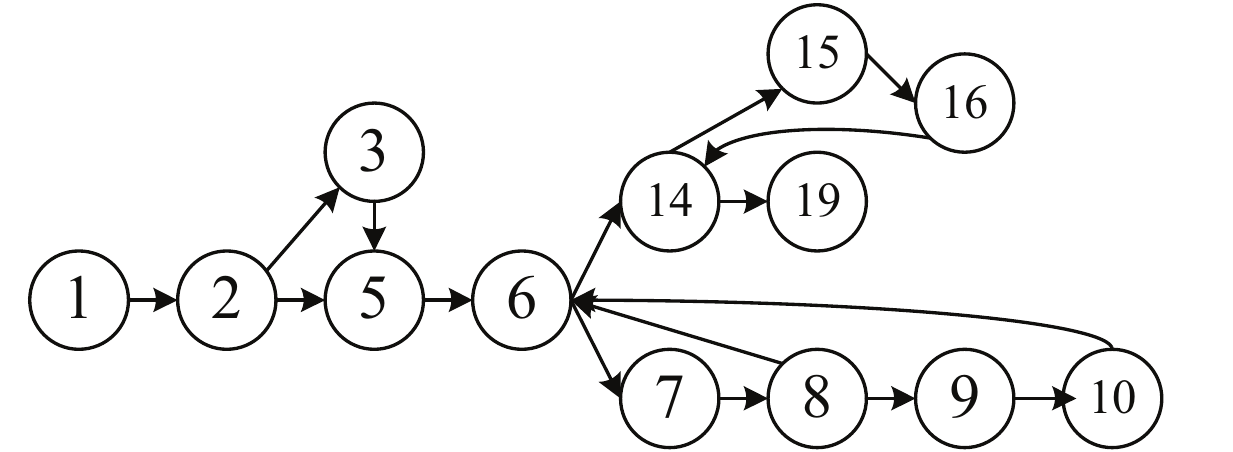}}
  \subfigure[Dominantor tree of the CFG]{
    \label{fig:domtree} 
    \includegraphics[width=0.8\columnwidth]{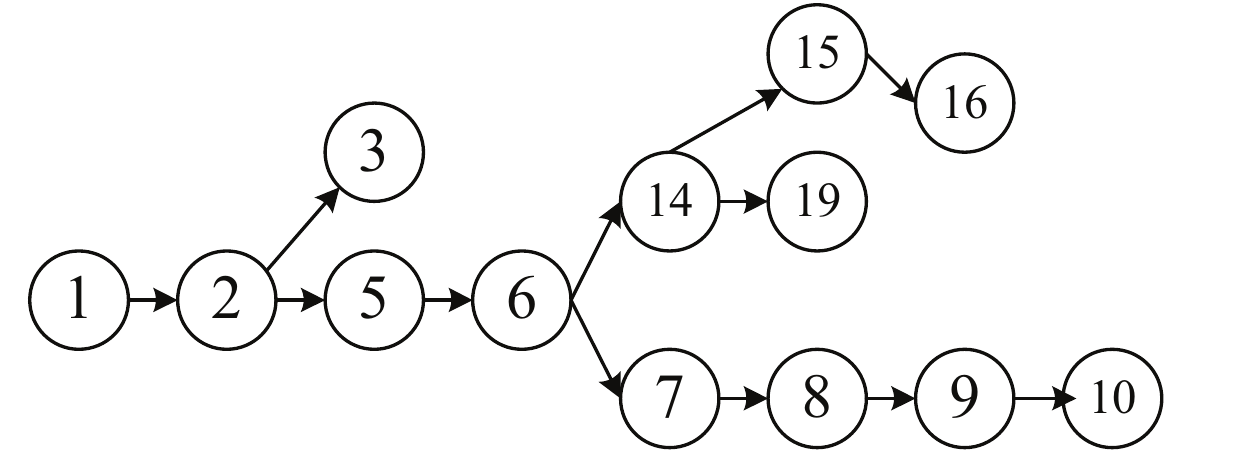}}
  \subfigure[Code splits]{
    \label{fig:slice} 
    \includegraphics[width=0.8\columnwidth]{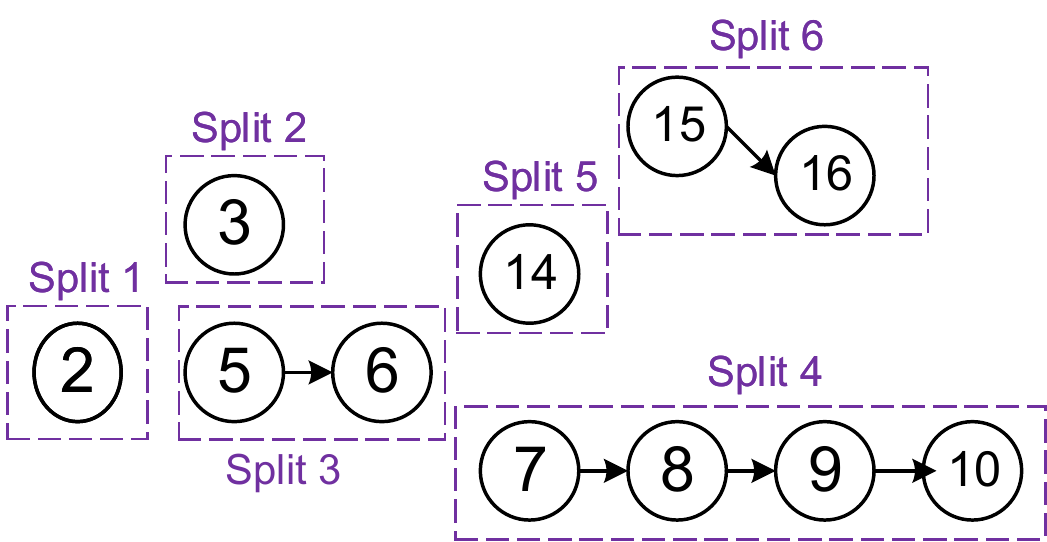}}
  \caption{Code splitting example}
  \label{fig:codeslicing} 
  \vspace{-13pt}
\end{figure}

Fig.~\ref{fig:codeslicing} illustrates the steps of code splitting. The CFG for the example in Fig.~\ref{fig:runningexample} is shown in Fig.~\ref{fig:cfg}. The resulting dominator tree is in Fig.~\ref{fig:domtree}. Excluding the start and end node, the dominator tree is segmented to six parts, as shown in Fig.~\ref{fig:slice}. We do not show the generated split AST for each split code due to page limits.

\begin{figure}[t]
\centering
\subfigure[Left split AST $t$]{
\includegraphics[width=0.83\columnwidth]{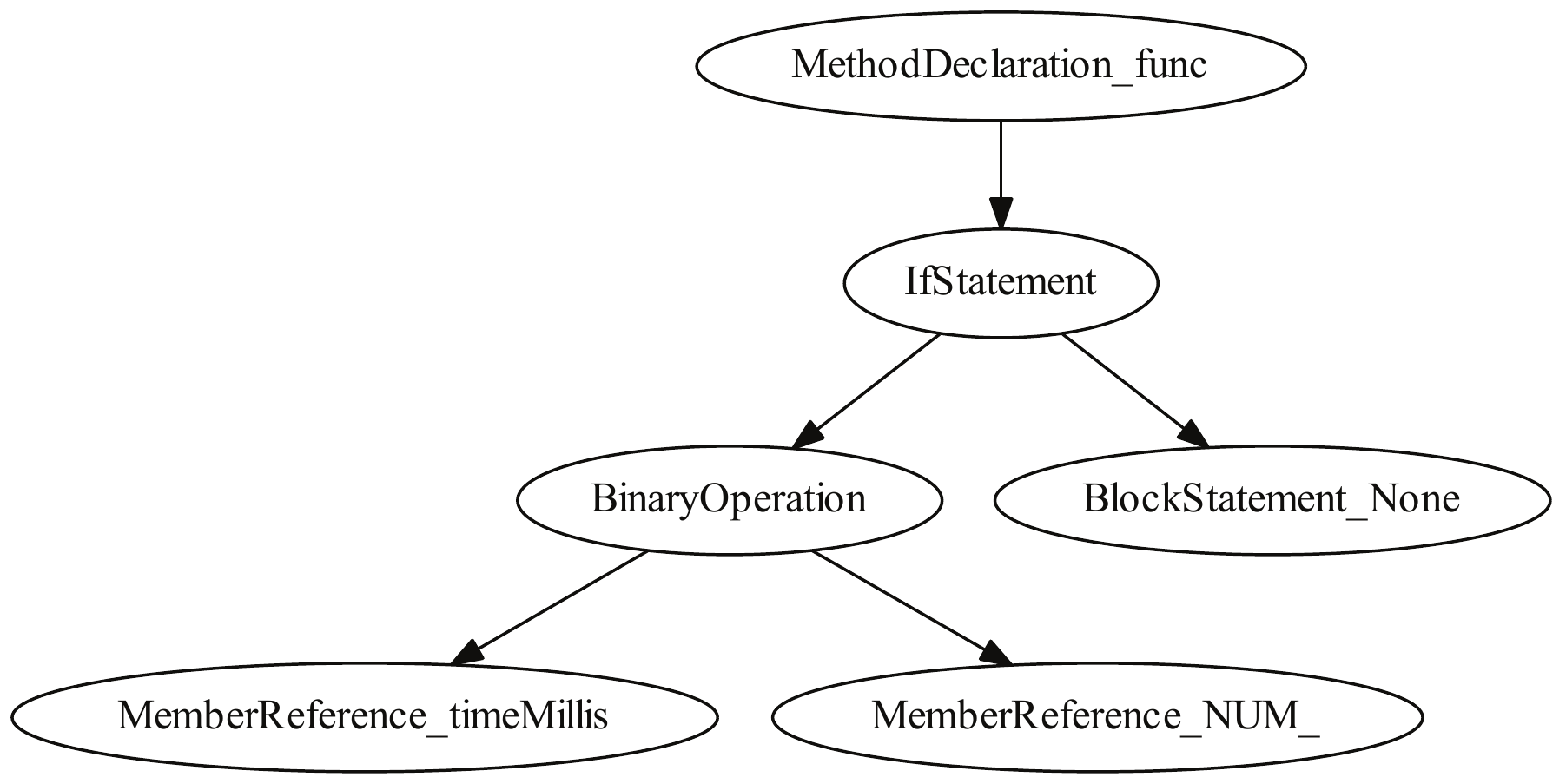}
}
\subfigure[Right split AST $t'$]{
\includegraphics[width=0.78\columnwidth]{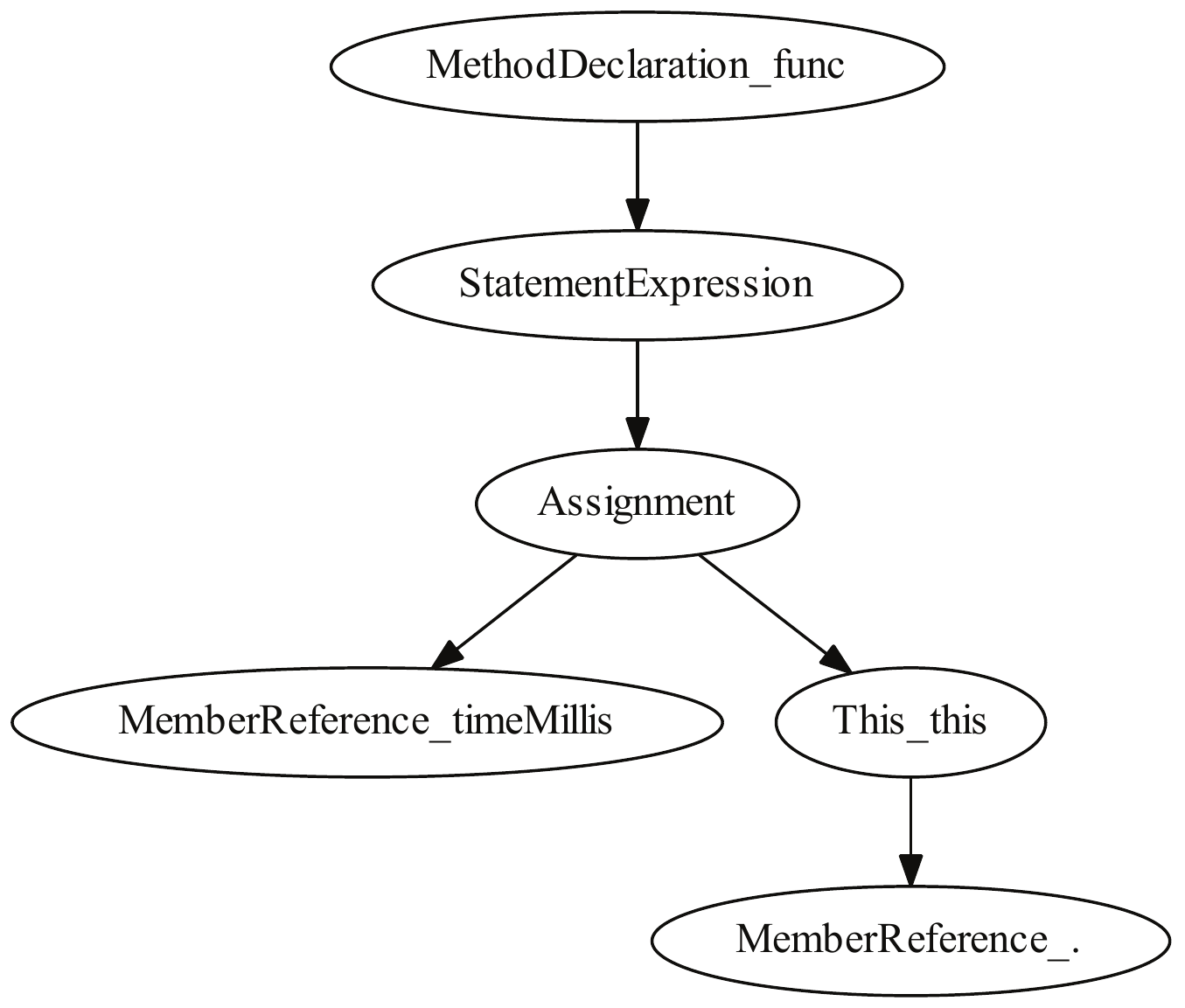}
}
\caption{An illustrative example of split ASTs}
\label{fig:sliced_ast}
\end{figure}

\begin{algorithm}[t]
\caption{The procedure to obtain syntax embedding for split AST $t$}
\label{alg:tree}
\small
\LinesNumbered
\KwIn{AST $t$ with node set $\Vcal(t)$}
\KwOut{Syntax embedding $\ebf_t$}
\SetKwFunction{FCall}{Tree-LSTM}
$v\leftarrow root(t)$\;
$\hbf_v \leftarrow$  Tree-LSTM($v$)\;
$\ebf_t\leftarrow \hbf_v$\;
 \SetKwProg{Fn}{Function}{:}{}
 \Fn{\FCall{$v$}}{
\If{$\Ccal_v\neq \emptyset$}
{
\For{$c\in\Ccal_v$}{$\hbf_c\leftarrow$ Tree-LSTM($c$)\;}
Compute $\hbf_v$ by Eq.~\ref{equ:TreeLSTM}\;}
\Else {
Randomly initiate $\hbf_c$\;
$\Ccal_v \leftarrow = \{c\} $\;
Compute $\hbf_v$ by Eq.~\ref{equ:TreeLSTM} \;}
\KwRet $\hbf_v$\;
  }
\end{algorithm}

\begin{figure*}[t]
\centering
\includegraphics[width=0.98\textwidth]{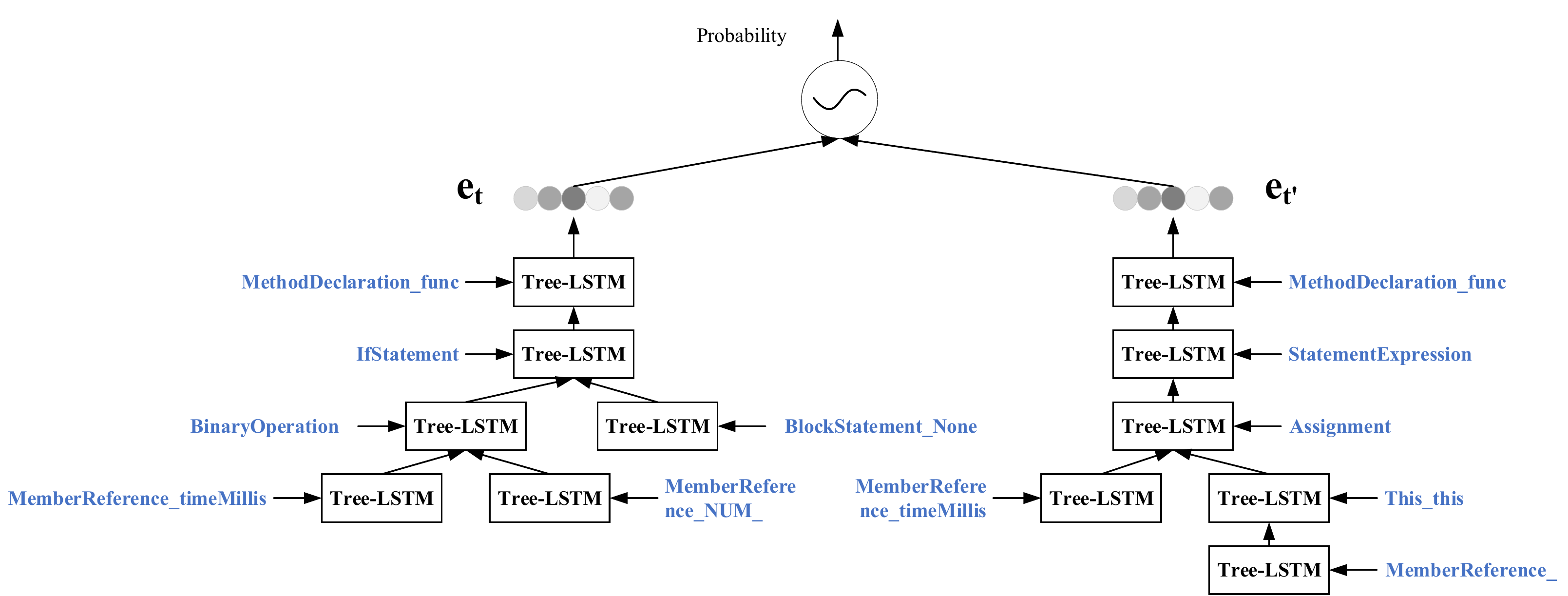}
\caption{An illustrative example of Syntax Embedding Pre-training (SEP) model on $t,t'$}
\label{fig:sep}
\end{figure*}

Note that, although there are various ways to split the code, 
we choose to leverage the blocks in the dominator tree~\cite{prosser1959applications} for two reasons:
\begin{enumerate}[leftmargin=13pt]
\item The consecutive statements inside each block typically represent a piece of business logic which contains richer information than single statement (more analysis is provided in Sec.~\ref{sec:contri}.).
\item There are no cyclic structures among blocks in the dominator tree and this greatly simplifies the design of the tree encoding phase (see the explanation of Eq.~\ref{equ:lossHID} in Sec.~\ref{sec:pretrain}).
\end{enumerate}

\section{Tree Encoding}\label{sec:pretrain}
The tree encoding component targets at each split AST $t$, to learn a vectorized numerical representation, i.e., \textit{syntax embedding} $\ebf_t\in \Rcal^{L}$, where $L$ is the embedding size. 
As mentioned in Sec.~\ref{sec:overview}, the syntax embedding is first pre-trained in a task-independent manner and then fine-tuned for the code summarization task. 
In this section, we introduce in detail the Syntax Embedding Pre-training (SEP) model for tree encoding in \ours.

The SEP model takes two split ASTs $t$ and $t'$ as inputs, where an AST $t$ has a set of nodes $\Vcal(t)$. Each node $v\in \Vcal(t)$ is assigned by a \textit{type\_value}
in the AST generation step, which is denoted as $\xbf_v$. For instance, the AST node with type \texttt{MemberReference} and value \texttt{timeMillis} will be assigned with \texttt{MemberReference\_timeMillis}. Each non-leaf node $v$ has a non-empty set of children nodes, which is defined as $\Ccal_v$. 

For each node $v$, we apply the Child-Sum Tree-LSTM~\cite{Sheng2015Improved} which outputs the hidden state $\hbf_v\in\Rcal^L$ based on the inputs including hidden states of children nodes, i.e., $\hbf_c$ ($c\in \Ccal_v$) and the embedding vector of its type\_value, i.e., $\xbf_v$. 
Given a non-leaf node $v$ where $\Ccal_v\neq\emptyset$, the Child-Sum Tree-LSTM operates as follows: 
\begin{equation} \label{equ:TreeLSTM}
\begin{aligned}
\tilde{\hbf}_{v} &= \sum_{c\in\Ccal_v}\hbf_c\\
\ibf_{v} &= \sigma \big(\Wbf^{i} \xbf_{v}  + \Ubf^i \tilde{\hbf}_{v}   + \bbf^i \big)\\
\fbf_{v,c} &= \sigma \big(\Wbf^{f} \xbf_{v}  + \Ubf^{f} \hbf_{c} + \bbf^f \big)\\
\obf_v &=\sigma \big(\Wbf^{o} \xbf_{v} + \Ubf^{o} \tilde{\hbf}_{v}  + \bbf^{o} \big) \\
\ubf_v &=tanh \big(\Wbf^{u} \xbf_{v} + \Ubf^{u} \tilde{\hbf}_{v}  + \bbf^{u} \big) \\
\mbf_v &=\ibf_v \odot \ubf_v + \sum_{c\in \Ccal_v} \fbf_{v,c} \odot \mbf_c \\
\hbf_{v} &= \obf_v \odot tanh(\mbf_v) 
\end{aligned}
\end{equation}
where $\ibf$, $\fbf$ and $\obf$ are the \emph{input}, \emph{forget}, \emph{output} gates, respectively. $\mbf$ is the \emph{memory cell}. $\bbf$ is the bias vector, $\Wbf$ and $\Ubf$ are weight matrices, $\sigma$ is the sigmoid activation function, $tanh$ is the tanh activation function, and $\xbf_{v}$ is the input embedding of the type\_value of node $v$.

For each split AST $t$, we use the output of the root node as the syntax embedding $\ebf_t$, which is computed according to Alg.~\ref{alg:tree}. Note that $\tilde{\hbf}_{v}$ in Eq.~\ref{equ:TreeLSTM} is the temporary hidden state summed over all children nodes $c\in\Ccal_v$. For a leaf node $v$ with $\Ccal_v=\emptyset$, a ``virtual" child node is randomly initialized to facilitate the computation of $\tilde{\hbf}_{v}$ (see lines 10-12 in Alg.~\ref{alg:tree}). 
Finally, the SEP model predicts $f(t,t')\in(0,1)$ based on the syntax embeddings $\ebf_t$ and $\ebf_{t'}$: 
\begin{equation}\label{equ:SEP}
f(t,t')=\sigma\big(concat(\ebf_t,\ebf_{t'})\big),
\end{equation}
where $\sigma$ is the sigmoid activation function, $concat(\cdot,\cdot)\in\Rcal^{2L}$ is the concatenation of $\ebf_t$ and $\ebf_{t'}$. 
In Fig.~\ref{fig:sliced_ast}, we illustrate two split ASTs, which are some of the split ASTs after splitting the example in Fig.~\ref{fig:runningexample}. Fig.~\ref{fig:sep} depicts the corresponding SEP model when learning these two split ASTs.

To learn the parameters, SEP optimizes the cross-entropy loss for next split AST prediction. 
\begin{equation}\label{equ:lossHID}
\small
\mathcal{L}_{pre} =\sum_{t,t'} \bigg( \hat{f}(t,t')\log f(t,t') + \big(1-\hat{f}(t,t')\big) \log \big(1-f(t,t')\big) \bigg),
\end{equation}
where $\hat{f}(t,t')$ is the gold standard label. If a split AST $t$ is proceeded by $t'$ in the dominator tree, i.e. $(t\rightarrow t')\in\Ecal(DT)$, then $\hat{f}(t,t')=1$, otherwise, $\hat{f}(t,t')=0$. 
Note that $\mathcal{L}_{pre}$ is the loss associated with source code, thus the pre-training is unsupervised and is less affected by the training data, e.g., the number of paired comments with source code and the quality of comments. It is worthy noting that our choice of code splitting based on dominator tree simplifies the design for predicting next split AST, since there is no cyclic structures after splitting. If the resulted code splits form a cycle graph (e.g., Split 1 is the subsequent block of Split 6 in Fig.~\ref{fig:slice}) after using other splitting methods, there will be an infinite number of $\langle \text{current split}, \text{next split}\rangle$ pairs. In this case, it will be necessary to define a strategy to limit the number of such pairs to be predicted in the pre-train task. Unfortunately, such a strategy is not easy to design, as any manual rule will make some pairs being learned less than others or hinder the information propagation along the cycle in the dependence learning. As a comparison, our approach explained in Sec.~\ref{sec:code_split} opts to split based on dominator tree and eliminates the possibility of the cyclic structure.

\section{Summarizing Code with Transformer }\label{sec:summarization}

\begin{figure}[t]
\begin{center}
\includegraphics[width=\columnwidth]{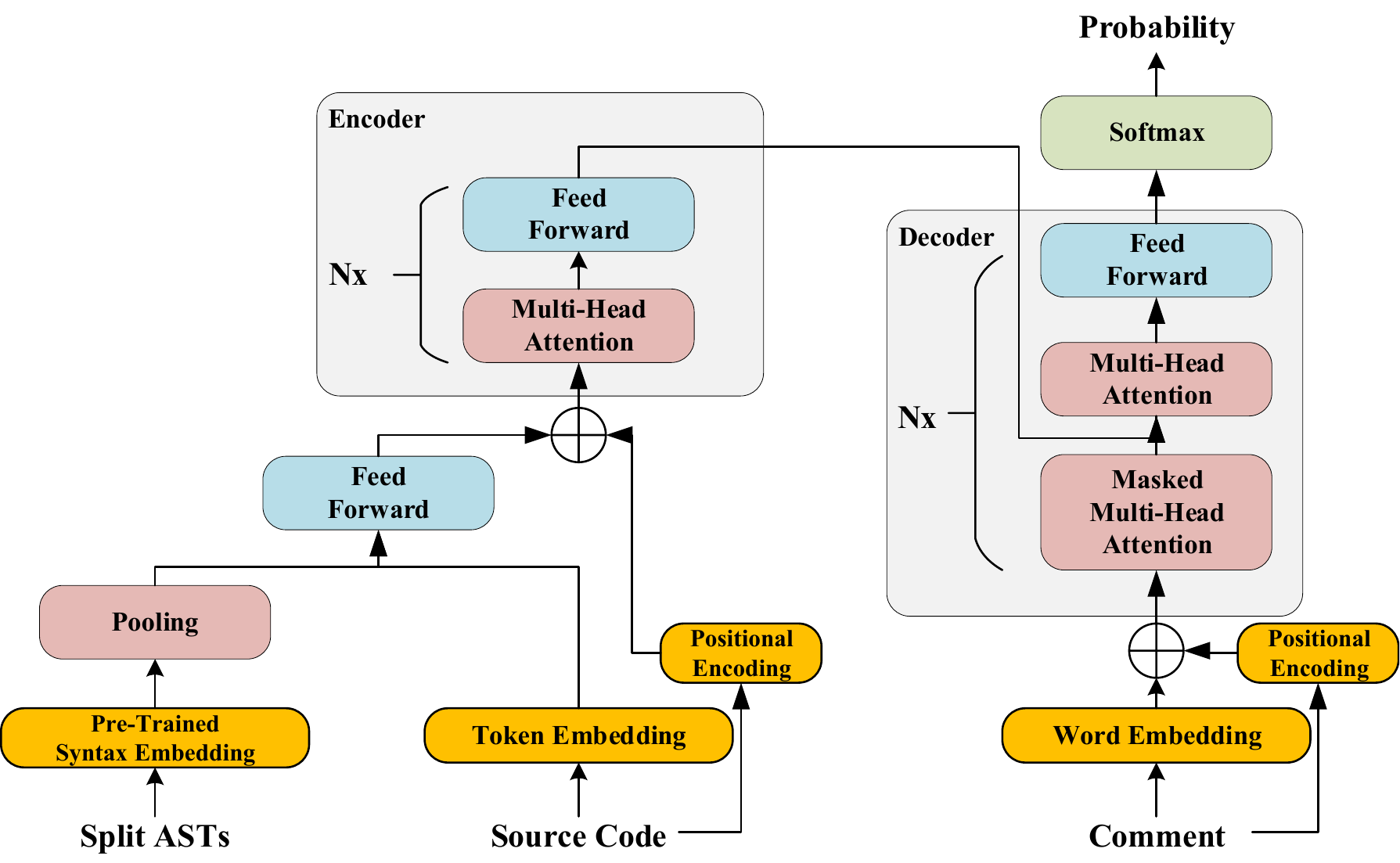}
\caption{Summarization model in \ours}
\label{fig:transformer}
\end{center}
\vspace{-15pt}
\end{figure}

The overall code summarization model in \ours, as shown in Fig.~\ref{fig:transformer}, follows an encoder-decoder framework~\cite{ChoMGBBSB14}, and we adopt Transformer~\cite{Vaswani2017Attention} as the implementation. 

\vspace{5pt}
\noindent\textbf{Input.} Recall that the $n$-th training sample is defined as $\langle \xbf^n, \ybf^n\rangle$. In the sequel, the superscript $n$ will be omitted when there is no ambiguity. Code splitting (Sec.~\ref{sec:code_split}) and Syntax Embedding Pre-training (Sec.~\ref{sec:pretrain}) obtain a set of $L$-dimensional vectors for the split AST, i.e., $\{\ebf_t|t\in \Tcal\}$, where $\ebf_t\in\Rcal^L$. 
The syntax embeddings for split ASTs flow through a pooling layer, which generates the representation of the overall syntax: 
\begin{equation}
\label{eq:avg}
AvgPooling(\ebf_1,\cdots,\ebf_{|\Tcal|})=[\frac{\sum_t \ebf_{t,1}}{|\Tcal|},\cdots,\frac{\sum_t \ebf_{t,L}}{|\Tcal |} ].
\end{equation} 
   
Each code token is also encoded by a numerical embedding vector. 
The syntax embeddings and code token embeddings are combined in a feed-forward layer. Note that each code embedding in a code sequence is concatenated with the same syntax embedding. 
Next, we will slightly abuse the notation here to use $\tilde{\mathbf{x}} \in \Rcal^{2L}$ to represented the concatenation of the output of the pooling layer and one token embedding. $\tilde{\mathbf{x}}$ is fed into a single-layer feedforward neural network:   
 \begin{equation}
FFN(\tilde{\mathbf{x}})=ReLU(\Wbf^F \tilde{\mathbf{x}}   +\bbf^F),
\end{equation}   
where $ReLU$ is the ReLU activation function. $\Wbf$ and $\bbf$ are weight matrix and vector, respectively. Intuitively, the output $FFN(\tilde{\mathbf{x}})$ is a weighted combination of syntax embedding and token embedding. This layer reveals that the understanding of source code incorporates both the syntax structure and the token sequence. 

To keep the relative position of a token in the code, position encodings are added to the input embeddings. The position encoding is formatted by:
\begin{equation}\label{equ:position}
pos(d,l)=\left\{
\begin{array}{rcl}
sin(d/10000^{l/L} )      &      & \textrm{if } l \textrm{ is even}\\
cos(d/10000^{l/L} )      &      & \textrm{if } l \textrm{ is odd}\\
\end{array} \right. 
\end{equation}
where $d$ is the token index in the source code, $l$ is the index of the embedding vector, and $L$ is the embedding size. $sin$ and $cos$ are the sin and cosine functions, respectively.

\vspace{5pt}
\noindent\textbf{Encoder.} 
The encoder is responsible for reading the input one token at a time and encoding the current information in a hidden state. 
The input embedding vector flows through a stack of encoder layers, where each encoder layer consists of a multi-head self-attention layer and a feedforward layer. 

The multi-head self-attention layer first constructs the input matrix $\mathbf{X}\in \Rcal^{D\times L}$ by stacking $D_n$ tokens and padding the rest $D-D_n$ rows with zeros. 
Three $L\times L$ trainable matrices $\Wbf^Q$, $\Wbf^K$ and $\Wbf^V$ transform the input matrix to \textit{queries} $\mathbf{Q}=\mathbf{X} \Wbf^Q$, \textit{keys} $\mathbf{K}=\mathbf{X} \Wbf^K$ and \textit{values} $\mathbf{V}=\mathbf{X} \Wbf^V$. The multi-head mechanism splits the \textit{queries}, \textit{keys} and \textit{values} to $H$ parts, each of $\mathbf{Q}^{h}$, $\mathbf{K}^{h}$ and $\mathbf{V}^{h}$ has the dimension size $D\times L_h$ so that each head can focus on different parts of the input. The output of each head is computed as:
\begin{equation}\label{equ:attention}
Attention(\mathbf{Q}^{h},\mathbf{K}^{h},\mathbf{V}^{h})=softmax(\frac{\mathbf{Q}^{h},{\mathbf{K}^{h}}^T}{\sqrt{L_h}})\mathbf{V}^{h}.
\end{equation}

The outputs of each head are concatenated and fed to a single-layer feedforward neural network. After that, each encoder layer also employs residual connection and normalization~\cite{Vaswani2017Attention}. 

\vspace{5pt}
\noindent\textbf{Decoder.} 
The decoder is responsible for generating a comment word based on the
previous output and the current encoding state. The decoder is
composed of a stack of identical layers. In each layer, there are
three sublayers. The first sublayer is a masked multi-head self-attention
mechanism. The second sublayer is a multi-head self-attention layer. 
The operation of the multi-head attention in the decoder is similar to Eq.~\ref{equ:attention}. 
The difference between masked multi-head self-attention and multi-head self-attention is that the former looks to earlier positions in the output. 
The third sublayer is a fully connected feedforward network. 
Finally, the output of the stacked layers is passed through a softmax layer to produce the probabilities of each candidate word. 

\vspace{5pt}
\noindent\textbf{Optimization.} The probabilities output by the decoder will be used in the standard cross-entropy loss which can be minimized by algorithms such as gradient descent to estimate the parameters of the overall summarization model.

\section{Experiment}\label{sec:experiment}

We conduct experiments to analyze the performance of \ours\footnote{Our implementation is available at \url{https://github.com/XMUDM/BASTS}.}. We focus on the following research questions:

\textbf{RQ1:} Does \ours generate better comments, compared with the state-of-the-art approaches?
 
\textbf{RQ2:} Does each component contribute to the performance of \ours?

\textbf{RQ3:} What are the impacts of dataset characteristics on the quality of generated comments?

\subsection{Experimental Setup}\label{sec:exp_setup}

\noindent\textbf{Data.} We use two public benchmarks: the Java dataset\footnote{\url{https://github.com/xing-hu/EMSE-DeepCom}}~\cite{Hu2019HDeep} which was collected from GitHub's Java repositories, and the Python dataset\footnote{\url{https://github.com/github/CodeSearchNet}}~\cite{abs-1909-09436} which is provided by the CodeSearchNet challenge. We replace number, string and boolean variables with special tokens $\langle$\texttt{NUM}$\rangle$, $\langle$\texttt{STR}$\rangle$ and $\langle$\texttt{BOOL}$\rangle$ in source code, respectively.
For both datasets, we use
SCITOOLS\footnote{\url{https://scitools.com/support/modifying-the-control-flow-graph}}
to generate CFGs. We also use the Javalang
library\footnote{\url{https://pypi.org/project/javalang}} and the ast
library\footnote{\url{https://docs.python.org/3/library/ast.html}} to generate
ASTs for Java dataset and Python dataset, respectively.  The source code in
Java is parsed into tokens by Javalang. Python dataset has been tokenized
already by its authors. For both datasets, identifiers are split
into subtokens (i.e., a variable \texttt{camelCase} yields two subtokens \texttt{camel} and
\texttt{case}), and comments are parsed into words by
NLTK\footnote{\url{https://www.nltk.org}}.

The statistics of datasets are shown in Tab.~\ref{tab:data}. For both datasets, we use the default 90/5/5 splits provided by the authors~\cite{Hu2019HDeep,abs-1909-09436} for training, validation and test sets. It is worth pointing out that, for Java dataset, there are 5,824 instance in the validation set and 5,341 instances in the test set which also exist in the training set. We remove them from the validation/test set. We also delete erroneous methods which can not generate CFGs or ASTs in two datasets.

\vspace{5pt}
\noindent\textbf{Hardware.} The experiments were run on a machine with two Intel(R) Xeon(R) CPU E5-2678 v3 @ 2.50GHz, 256 GB main memory and a GeForce RTX 2080 Ti graphics card. 

\vspace{5pt}
\noindent\textbf{Evaluation Metrics.} We use BLEU, METEOR and ROUGE which are widely used in previous works for code summarization~\cite{abs-2010-01410}. 
BLEU basically computes the averaged percentage of n-gram matches over each $n=1,\cdots,N$:
\begin{equation}
P_n=\frac{\sum_{gram_n \in \mathbf{s}} \#{match}(gram_n) } {  \#(gram_n) },\,\,\,\,BLEU=\rho (\prod_n P_n)^{1/N}\nonumber
\end{equation}
where $\sum_{gram_n \in \mathbf{s}} \#{match}(gram_n) $ is the number of matching n-grams in the result and the gold standard $\mathbf{s}$, $ \#(gram_n) $ is the number of n-grams in the result, $\rho$ is a brevity penalty. 
We adopt S-BLEU and C-BLEU, which indicate average sentence-level BLEU score and corpus-level average BLEU score, respectively. 

ROUGE compares the generated comment with the gold-standard comment by counting the number of their overlapping textual units. We consider textual units such as unigram, bigram and longest common sequence (LCS), i.e., ROUGE-1, ROUGE-2 and ROUGE-L. 
We adopt F-Score for detailed calculation.
For example, ROUGE-L F-Score is computed as:
\begin{equation}
R=\frac{\#LCS(\mathbf{s},\ybf)}{m},\,\,\,\,P=\frac{\#LCS(\mathbf{s},\ybf)}{n},\,\,\,\,F= \frac{2RP}{R+P}\nonumber
\end{equation} 
where $\#LCS(\mathbf{s},\ybf)$ is the number of matching longest common sequence between the gold-standard and the result summary, $m$ is the number of LCSs in the result comment and $n$ is the number of LCSs in the gold standard summary. The computation of ROUGE-1 and ROUGE-2 is similar by replacing LCS with matching unigrams or bigrams. 

METEOR\footnote{\href{http://www.cs.cmu.edu/~alavie/METEOR/}{http://www.cs.cmu.edu/$\sim$alavie/METEOR/}} scores the quality of translation by aligning them to one or more gold-standards. Alignments are based on exact, stem, synonym, and paraphrase matches between words and phrases. The metric includes several free parameters that are tuned to emulate human judgments. 

\begin{table}[t]
\caption{Statistics of datasets}
\begin{center}
\begin{tabular}{|c|c|c|c|c|}
\hline
\textbf{Dataset} & \textbf{\#Methods} & \textbf{\#Tokens} & \textbf{\#Words} & \textbf{\#Split ASTs} \\ \hline\hline
Java & 441,487 & 45,790  & 55,015 & 1,273,301 \\ \hline
Python & 241,322 & 237,939 & 75,883 & 1,033,210 \\ \hline
\end{tabular}
\end{center}
\label{tab:data}
\vspace{-10pt}
\end{table}

\begin{table*}[t]
\caption{Comparative code summarization performance on Java dataset. Best performance are in bold font, and improvement percentage of \ours w.r.t the best baseline are in brackets.}
\begin{center}
\begin{tabular}{|c|c|c|c|c|c|c|}
\hline
\textbf{Method}                                                 & \textbf{S-BLEU}                                                               & \textbf{C-BLEU}                                                               & \textbf{METEOR}                                                               & \textbf{ROUGE-1}                                                              & \textbf{ROUGE-2}                                                              & \textbf{ROUGE-L}                                                              \\ \hline\hline
CODE-NN                                                         & 24.22                                                                         & 14.68                                                                         & 14.79                                                                         & 34.60                                                                         & 17.85                                                                         & 33.18                                                                         \\ \hline
Hybrid-DRL                                                      & 20.16                                                                         & 10.99                                                                         & 13.15                                                                         & 30.45                                                                         & 13.80                                                                         & 29.22                                                                         \\ \hline
Hybrid-DeepCom                                                  & 29.19                                                                         & 19.82                                                                         & 18.21                                                                         & 40.52                                                                         & 23.46                                                                         & 38.70                                                                         \\ \hline
AST-attendgru                                                   & 27.50                                                                         & 17.85                                                                         & 16.44                                                                         & 36.73                                                                         & 20.83                                                                         & 35.32                                                                         \\ \hline
\begin{tabular}[c]{@{}c@{}}ASTNN\\ (batch size=64)\end{tabular} & 25.40                                                                         & 15.94                                                                         & 16.92                                                                         & 38.21                                                                         & 19.66                                                                         & 36.10                                                                         \\ \hline
Transformer                                                     & 30.98                                                                         & 23.63                                                                         & 20.30                                                                         & 43.49                                                                         & 27.60                                                                         & 42.20                                                                         \\ \hline
Rencos                                                          & 32.90                                                                         & 26.03                                                                         & 21.52                                                                         & 44.12                                                                         & 29.31                                                                         & 42.84                                                                         \\ \hline
\ours                                                           & \begin{tabular}[c]{@{}c@{}}\textbf{36.38}\\ ($\uparrow 10.58\%$)\end{tabular} & \begin{tabular}[c]{@{}c@{}}\textbf{30.37}\\ ($\uparrow 16.67\%$)\end{tabular} & \begin{tabular}[c]{@{}c@{}}\textbf{24.12}\\ ($\uparrow 12.08\%$)\end{tabular} & \begin{tabular}[c]{@{}c@{}}\textbf{49.63}\\ ($\uparrow 12.49\%$)\end{tabular} & \begin{tabular}[c]{@{}c@{}}\textbf{33.13}\\ ($\uparrow 13.03\%$)\end{tabular} & \begin{tabular}[c]{@{}c@{}}\textbf{47.85}\\ ($\uparrow 11.69\%$)\end{tabular} \\ \hline
\end{tabular}
\end{center}
\label{tab:JavaLarge}
\vspace{-10pt}
\end{table*}

\begin{table*}[t]
\caption{Comparative code summarization performance on Python dataset. Best performance are in bold font, and improvement percentage of \ours w.r.t the best baseline are in brackets.}
\begin{center}
\begin{tabular}{|c|c|c|c|c|c|c|}
\hline
\textbf{Method} & \textbf{S-BLEU}                                                              & \textbf{C-BLEU}                                                       & \textbf{METEOR}                                                              & \textbf{ROUGE-1}                                                              & \textbf{ROUGE-2}                                                             & \textbf{ROUGE-L}                                                              \\ \hline\hline
CODE-NN         & 15.37                                                                        & 1.68                                                                  & 7.25                                                                         & 18.57                                                                         & 3.28                                                                         & 17.02                                                                         \\ \hline
Hybrid-DRL      & 14.09                                                                        & 2.25                                                                  & 7.80                                                                         & 18.96                                                                         & 4.18                                                                         & 17.77                                                                         \\ \hline
Hybrid-DeepCom  & 10.24                                                                        & 0.85                                                                  & 5.03                                                                         & 13.35                                                                         & 2.36                                                                         & 12.41                                                                         \\ \hline
AST-attendgru   & 14.47                                                                        & 1.17                                                                  & 6.42                                                                         & 17.05                                                                         & 2.85                                                                         & 15.76                                                                         \\ \hline
Transformer     & 14.80                                                                        & 1.64                                                                  & 7.60                                                                         & 19.30                                                                         & 4.00                                                                         & 17.91                                                                         \\ \hline
Rencos          & 15.03                                                                        & \textbf{3.31}                                                         & 8.73                                                                         & 20.90                                                                         & 5.51                                                                         & 19.40                                                                         \\ \hline
\ours           & \begin{tabular}[c]{@{}c@{}}\textbf{15.97}\\ ($\uparrow 6.25\%$)\end{tabular} & \begin{tabular}[c]{@{}c@{}}2.77\\ ($\downarrow 16.31\%$)\end{tabular} & \begin{tabular}[c]{@{}c@{}}\textbf{9.79}\\ ($\uparrow 12.14\%$)\end{tabular} & \begin{tabular}[c]{@{}c@{}}\textbf{24.61}\\ ($\uparrow 17.75\%$)\end{tabular} & \begin{tabular}[c]{@{}c@{}}\textbf{6.18}\\ ($\uparrow 12.16\%$)\end{tabular} & \begin{tabular}[c]{@{}c@{}}\textbf{22.72}\\ ($\uparrow 17.11\%$)\end{tabular} \\ \hline
\end{tabular}
\end{center}
\label{tab:Python}
\vspace{-10pt}
\end{table*}

\subsection{Code Summarization Performance (RQ1)}
\label{sec:rq1}

To answer RQ1, we compare \ours with seven state-of-the-art methods:
\begin{enumerate}[leftmargin=13pt]
\item \textbf{CODE-NN}\footnote{\url{https://github.com/sriniiyer/codenn}}~\cite{Iyer2016Summarizing} uses LSTM  and a global attention mechanism in the seq2seq framework.    
\item \textbf{Hybrid-DRL}\footnote{\href{https://github.com/wanyao1992/code_summarization_public}{https://github.com/wanyao1992/code\_summarization\_public}}~\cite{Wan2018Improving} applies an actor-critic network. The actor network produces the hidden space. The critic network adopts BLEU as the reward function. 
\item \textbf{Hybrid-DeepCom}\footnote{\url{https://github.com/xing-hu/EMSE-DeepCom}}~\cite{Hu2019HDeep} uses GRU to build a code encoder for code tokens, and an AST encoder for serialized ASTs from SBT. 
A hybrid attention component fuses the lexical and syntactical information. 
\item \textbf{AST-attendgru}\footnote{\url{https://bit.ly/2MLSxFg}}~\cite{LeClairJM19} applies two GRU encoders on code sequences and serialized ASTs from SBT. A different attention mechanism compared to Hybrid-DeepCom is used for information fusion. 
\item \textbf{ASTNN}\footnote{\url{https://github.com/zhangj111/astnn}}~\cite{ZhangWZ0WL19} splits an AST into a sequence of statement trees. Statement trees are encoded by an RvNN based encoder. The learned sequence of statement tree representations are fed into GRU to generate the representation of the original code for code classification and clone detection. For a fair comparison, we feed the code representation learned by ASTNN into Eq.~\ref{eq:avg} for code summarization, and other components are the same as \ours.
\item \textbf{Transformer}\footnote{\url{https://github.com/wasiahmad/NeuralCodeSum}}~\cite{AhmadCRC20} employs the Transformer model~\cite{Vaswani2017Attention} in code summarization task. 
\item \textbf{Rencos}\footnote{\url{https://github.com/zhangj111/rencos}}~\cite{ZhangW00020} retrieves two code snippets that are most similar to the input code snippet. Then, it encodes the input and two retrieved code snippets, and generates the summary by fusing them during decoding.
\end{enumerate}

We use the default hyper-parameter settings provided by each method. Following Hu et al.~\cite{Hu2019HDeep}, we set maximal code length to $100$, maximal comment length to $30$, and maximal epochs to $50$ for all methods (we also report the impact of different code/comment lengths in Sec.~\ref{sec:rq3}). 
We use 512 as the embedding size and 128 as the batch size. The exception is that we use the batch size 64 for ASTNN since ASTNN with the batch size 128 will encounter out-of-memory error. For a fair comparison, we also compare ASTNN with \ours when both methods are trained with the batch size 64 in Sec.~\ref{sec:contri}. We use 8 heads in Transformer and \ours.
We adopt greedy search for AST-attendgru and Hybrid-DRL, following their original design. We use beam search with beam size $5$ for CODE-NN and Hybrid-DeepCom, as mentioned in their papers. For \ours, greedy search is used for simplicity. We adopt adam~\cite{KingmaB14} for optimization.

Tab.~\ref{tab:JavaLarge} and Tab.~\ref{tab:Python} report the results on two datasets. Note that the authors of ASTNN do not provide the implementation for splitting python code using their method. Hence we only evaluate ASTNN on Java dataset. From the results, we can observe that \ours is robustly better than state-of-the-art competitors. In most cases, \ours exceeds baselines by more than 10\% in terms of all measures. As pointed out by Gros et al.~\cite{abs-2010-01410}, existing neural code summarization methods may not indeed surpass traditional retrieval-based methods. Therefore, we compare \ours with Rencos which combines the merits of both retrieval-based and neural network-based methods. From Tab.~\ref{tab:JavaLarge} and Tab.~\ref{tab:Python}, we can see that \ours has better performance than Rencos in almost all cases, showing that \ours is better than the retrieval-based model.

\subsection{Contribution of Different Components of \ours (RQ2)}
\label{sec:contri}

\begin{figure*}[t]
\centering
\includegraphics[width=0.98\textwidth]{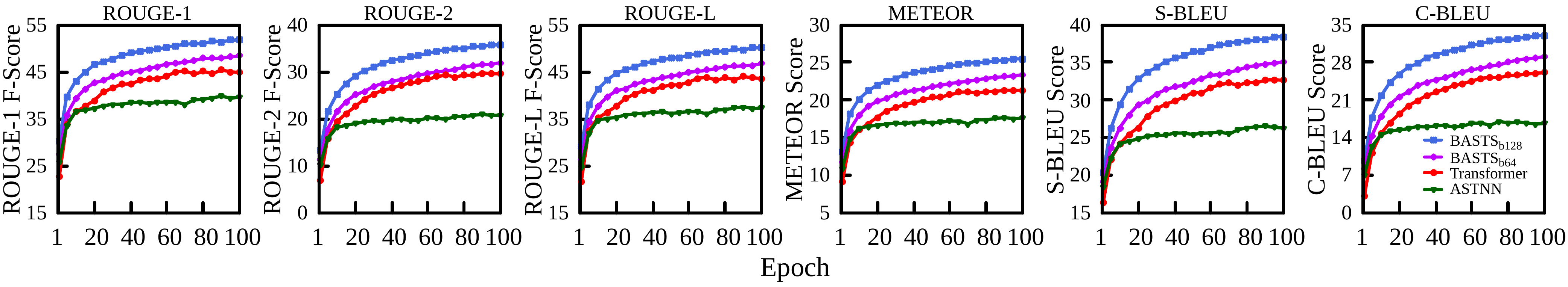}
\caption{Performance of different methods at each epoch on Java dataset.}
\label{fig:convergencelarge}
\end{figure*}

\begin{figure*}[t]
\centering
\includegraphics[width=0.98\textwidth]{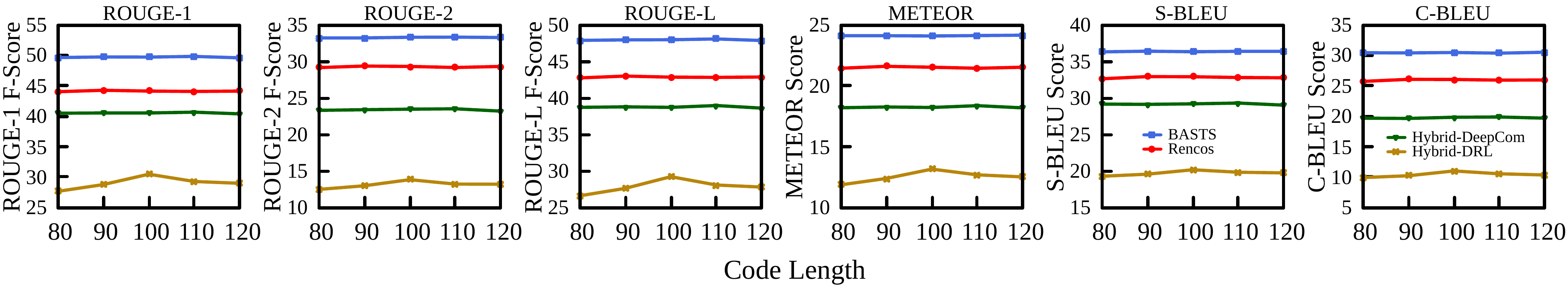}
\caption{Performance w.r.t varying code length on Java dataset.}
\label{fig:codelength}
\end{figure*}

\begin{figure*}[t]
\centering
\includegraphics[width=0.98\textwidth]{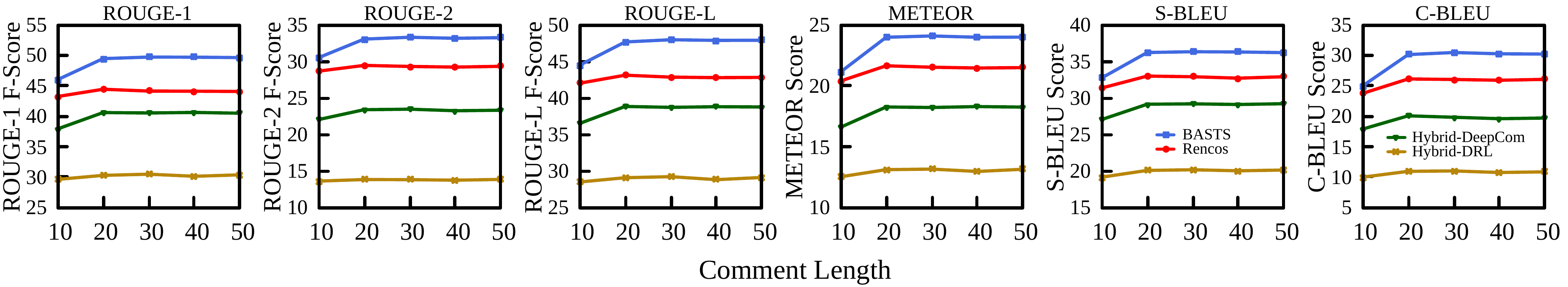}
\caption{Performance w.r.t varying comment length on Java dataset.}
\label{fig:commentlength}
\vspace{-6pt}
\end{figure*}

\begin{figure}[t]
\centering
\includegraphics[width=0.98\columnwidth]{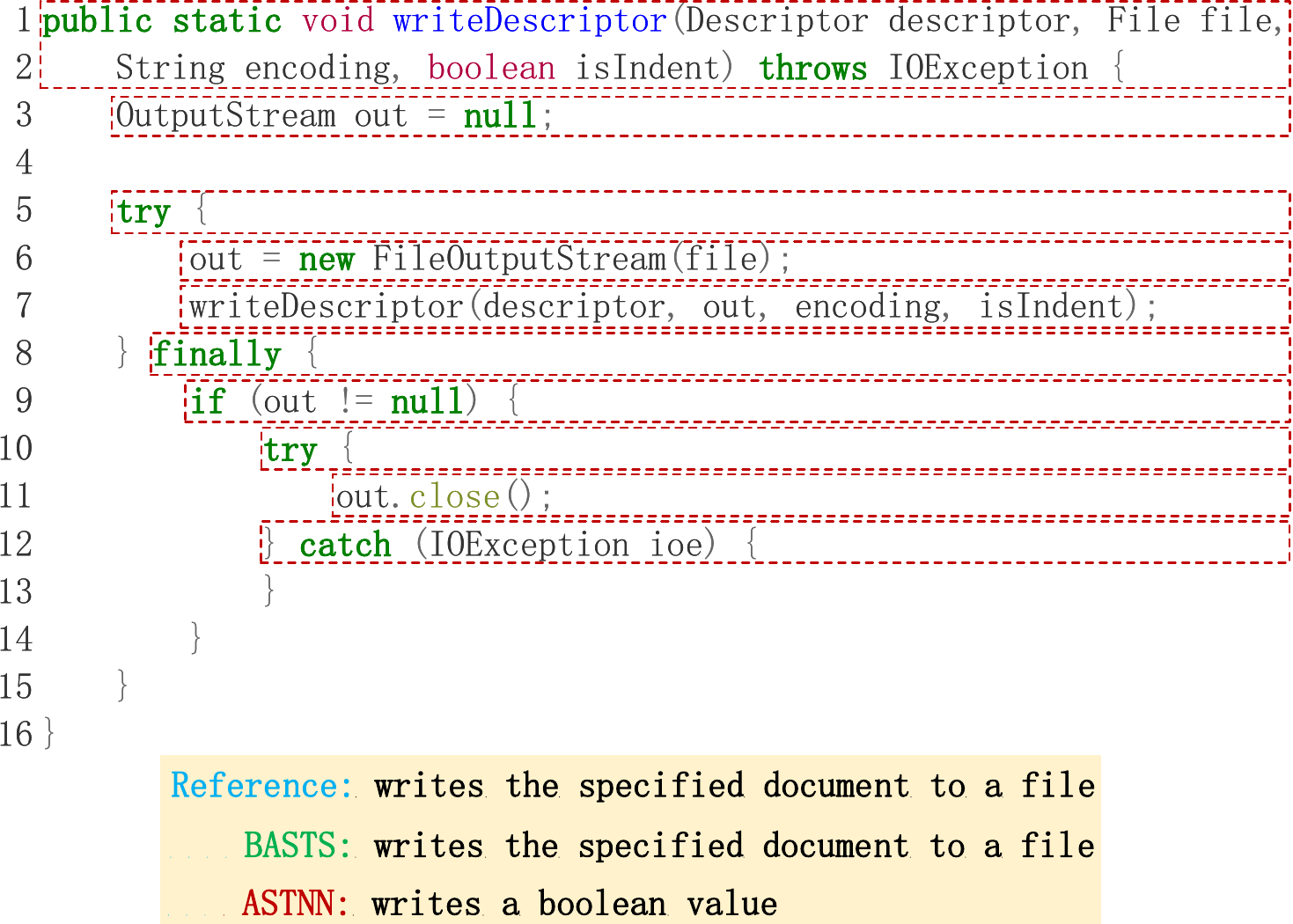}
\caption{Case Study: A comparison between ASTNN and \ours for test sample $\#3797$ in Java dataset. Red dashed lines indicate the statements decomposed by ASTNN.}
\label{fig:case_study}
\vspace{-5pt}
\end{figure}

To answer RQ2, we further investigate Transformer, ASTNN and \ours, which share some components in their models.
We report the performance of \oursnormal (batch size=128, i.e., \ours), \ourssmallbs (batch size=64), Transformer (batch size=128) and ASTNN (batch size=64) at each epoch
($i=1,\cdots,100$) on Java dataset in Fig.~\ref{fig:convergencelarge}. Due to page limit, we omit the similar results on Python dataset. 

Recently, the Transformer model has achieved a great success in various domains~\cite{abs-2009-06732}. To verify whether the superiority of \ours over existing methods is mainly caused by the use of the Transformer architecture, we compare \ours with the baseline Transformer~\cite{AhmadCRC20} which adopts the pure Transformer architecture for code summarization in Tab.~\ref{tab:JavaLarge}, Tab.~\ref{tab:Python} and Fig.~\ref{fig:convergencelarge}. From the results, we can see that \ours has much better performance than the pure Transformer architecture in code summarization task, showing that components like code splitting and pre-training in \ours indeed help improve its performance. Thus, we can conclude that using a more advanced neural architecture is not the only reason that \ours outperforms existing methods.

The next question is whether the way that \ours splits ASTs is better than the
existing AST splitting method, i.e., split ASTs into statement trees like ASTNN. Recall
that the difference between ASTNN and \ours in our experiments is how they
split ASTs, and other components in these two methods are the same. In the
main experiments in Tab.~\ref{tab:JavaLarge} and Tab.~\ref{tab:Python}, the
results of ASTNN are obtained when it uses the batch size 64 to avoid the
out-of-memory error. If we compare ASTNN and \ourssmallbs which both use the batch size 64 in
Fig.~\ref{fig:convergencelarge}, we can
observe that, after around 10 epochs, \ourssmallbs starts to significantly
outperform ASTNN. We further conduct case studies on the generated comments from ASTNN and \ours, and the statements decomposed by ASTNN. We show one case study from Java dataset in Fig.~\ref{fig:case_study}. From Fig.~\ref{fig:case_study}, we can see that the inadequate summary generated by ASTNN is related to the first statement only. The reason is that each statement tree split by ASTNN only consists of one statement, and ASTNN cannot capture the information beyond the statement boundary using its RvNN encoder only. If the sequential neural model (ASTNN uses GRU) which is used to model the sequence of statements does not fuse statement representations well, the generated comment will only include local information from one statement. \ours generates a perfect comment in Fig.~\ref{fig:case_study}. Assume that \ours also encounters the problem that its sequential model does not fuse the representations of split ASTs well like ASTNN in Fig.~\ref{fig:case_study}. Then, for such cases, \ours will still generate a comment containing more information than a single statement since split ASTs in \ours is based on blocks in the dominator tree (one block consists of several statements). From above observations and analysis, we can conclude that splitting ASTs based on the blocks in the dominator tree like \ours is better than splitting ASTs by statements like ASTNN.

\begin{figure}[t]
    \centering
    \begin{minipage}{0.48\columnwidth}
        \centering
        \vspace{-10pt}
        \includegraphics[width=1\textwidth]{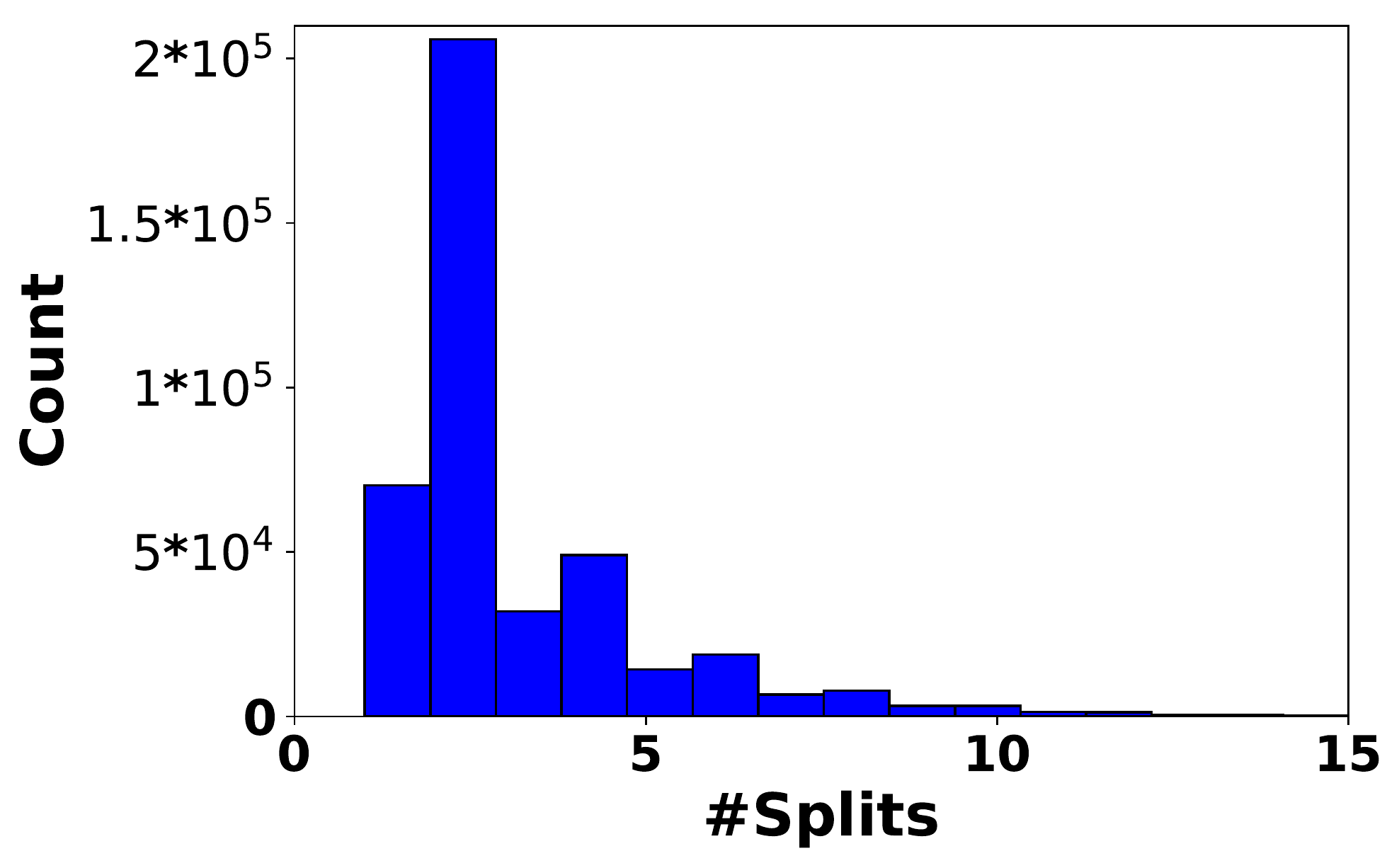} 
        \caption{Distribution of the number of split ASTs in Java dataset.}
		\label{fig:distribution}
    \end{minipage}\hfill
    \begin{minipage}{0.48\columnwidth}
        \centering
        \includegraphics[width=1\textwidth]{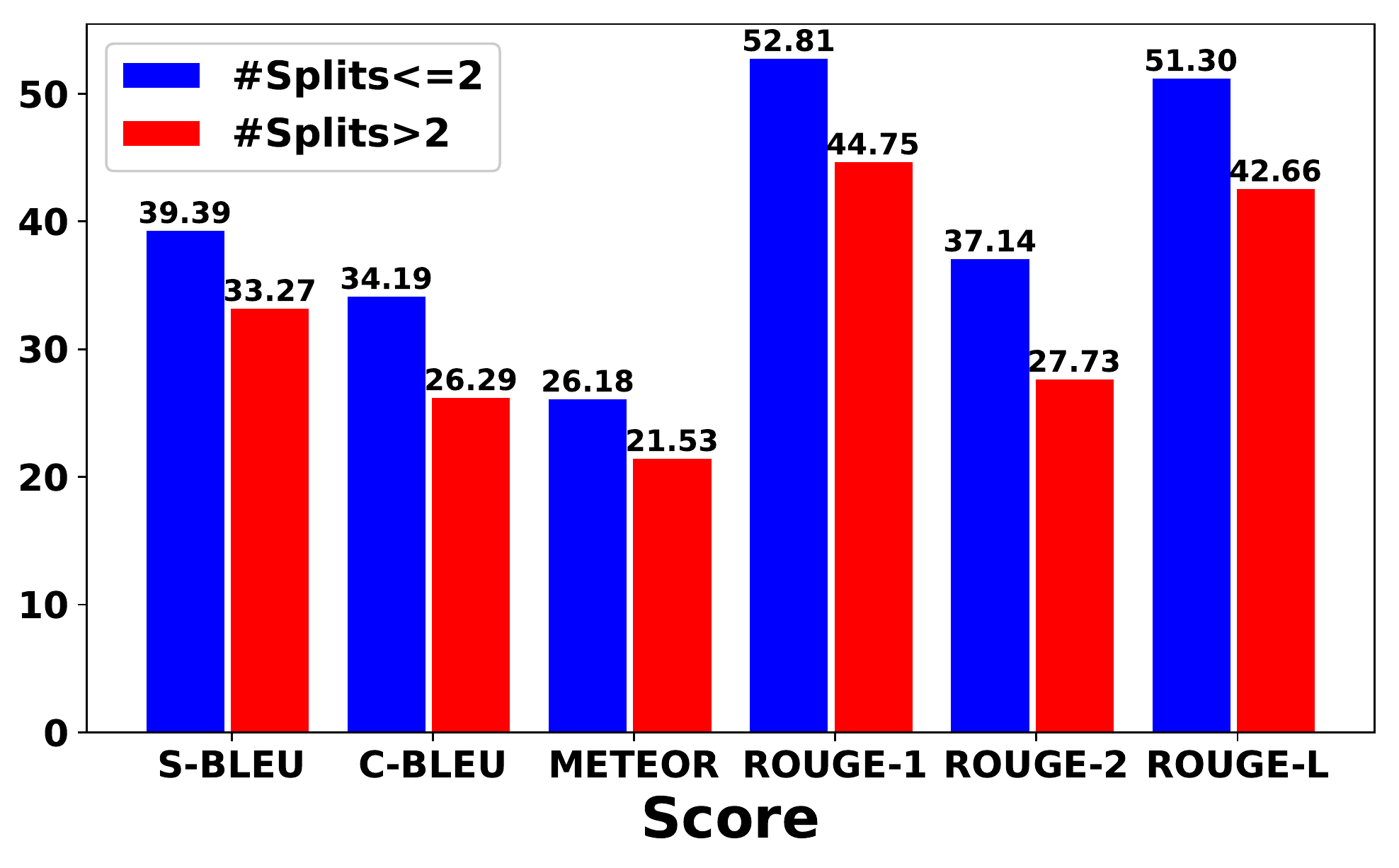} 
        \caption{Results on two partitions with different number of split ASTs in Java dataset.}
		\label{fig:resultslice}
    \end{minipage}
    \vspace{-15pt}
\end{figure}

\subsection{Impact of Dataset Characteristics (RQ3)}\label{sec:rq3}
Following Wan et al.~\cite{Wan2018Improving}, we analyze the impact of code
length and comment length of the Java data on the quality of generated
comments since the code length may affect the representation learning and the
comment length may affect the evaluation of comment generation. Due to space
limit, we do not show the similar analysis of Python dataset.

As shown in Fig.~\ref{fig:codelength} and Fig.~\ref{fig:commentlength}, \ours has best performance w.r.t. varying values of code and comment length on Java dataset, compared to baselines Rencos, Hybrid-DeepCom and Hybrid-DRL. From Fig.~\ref{fig:codelength} and Fig.~\ref{fig:commentlength}, we can see that \ours performs best when code length is $100$ or comment length is $30$. The underlying reason is that in Java dataset, $83\%$ of the methods are with code length $\leq 100$ , and $99.99\%$ of the comments are with length $\leq 30$. 
  
We next visualize the distribution of split ASTs by \ours in Java dataset in
Fig.~\ref{fig:distribution}.  We observe that most methods can be segmented to
two splits.  To investigate whether the number of split ASTs affects \ours's
performance, we divide the Java dataset into two partitions, one consisting of
instances that can be segmented to at most two splits and the other consisting
of instances that can be segmented to more than two splits.  Note that the
partitioning is performed on both training and test sets.   
Fig.~\ref{fig:resultslice} reports the performance of \ours on the two partitions. 
We can see that \ours performs slightly worse on the partition with more splits,  showing that it is more difficult to summarize such complex code (with more split ASTs). Observing both
Fig.~\ref{fig:resultslice} and Tab.~\ref{tab:JavaLarge}, we can see that
training \ours with complex code only (i.e., partition with \#splits$>$2) still give
better performance than Rencos (which is the best competitor as shown in
Tab.~\ref{tab:JavaLarge}) trained with all code in 4 out of 6 evaluation
metrics, which further demonstrates the superiority of \ours over existing
methods.

\section{Related Work}
\label{sec:rel}

\subsection{Source Code Summarization}
Source code comments are important for facilitating collaboration between
software developers. Much effort has been devoted to designing automatic code
summarization techniques in the past decade. Conventional approaches mostly
adopt Information Retrieval (IR) techniques~\cite{HaiducAMM10,SaltonWY75,DeerwesterDLFH90,HaiducAM10,RodegheroMMBD14,Aizawa03,EddyRKC13,MimnoLM07,Movshovitz-AttiasC13,FowkesCRALS16,FowkesCRALS17,BleiNJ01}
or methods from NLP
domain~\cite{SridharaHMPV10,SridharaPV11Generating,SridharaPV11Automatically,McBurneyM14,WangPV17}.

Recent breakthroughs of deep learning has significantly affect the design of automatic code summarization methods~\cite{abs-1909-04352}.
Most neural code summarization methods leverage the seq2seq framework in machine translation~\cite{ChoMGBBSB14}.  
Early approaches leverage various deep learning techniques like LSTM~\cite{Iyer2016Summarizing}, convolutional attention network~\cite{AllamanisPS16}, dual training~\cite{Wei2019Code} or multiple encoders~\cite{HuLXLLJ18} to generate code summarization based on input code tokens and the seq2seq framework.

Recently, there is a surge of works on using ASTs of the source code to guide the generation of summarization. Most of them firstly serialize ASTs into sequences.
One representative traversal method is the Structure-Based Traversal method
(SBT)~\cite{Hu2018Deep}. Given the serialized ASTs, existing works adopt sequential neural networks like
LSTM~\cite{Hu2018Deep}, GRU~\cite{Hu2019HDeep,LeClairJM19} or
Transformer~\cite{AhmadCRC20} to extract features for generating code
summarization. The limitation of these methods is that structural information
cannot be fully captured from serialized ASTs. To overcome this problem, some
works consider replacing the RNN based encoder with a Graph Neural Network
(GNN) based encoder~\cite{LeClairHWM20}
or feeding the outputs from the sequential encoder as the initial node
representations into the GNN based encoder~\cite{FernandesAB19}. 
These methods typically encounter high
overhead and face difficulties in training, as ASTs are deep with many nodes which are hard and slow to learn. 
In order to balance the
retention of structural information and the training difficulties, Wan et
al.~\cite{Wan2018Improving} transform ASTs into binary trees for efficient
training at the cost of partial information loss. Besides above methods of modeling ASTs, another way is to
represent a code snippet as the set of compositional paths in its AST for code
summarization~\cite{Alon2019code2seq}.

There are other directions for improving code summarization.
Wan et al.~\cite{Wan2018Improving} argue that the encoder-decoder framework suffers
from exposure bias, 
and they opt to deep
reinforcement learning for better code summarization. Chen and Zhou~\cite{ChenZ18} consider modeling code retrieval and code summarization tasks together.
Zhang et al.~\cite{ZhangW00020} design a new method which combines merits of both IR-based and neural network-based code summarization methods. Haque et al.~\cite{HaqueLWM20} add an additional RNN encoder to the seq2seq framework to encode file context and enhance code summarization. Gros et al.~\cite{abs-2010-01410} reviews the evaluation metrics used in existing code summarization works.

\subsection{Representation Learning for Programs}
The large volume of public available source code in source code management
platforms (e.g., GitHub) makes it possible to extract and learn knowledge from
code corpora. Hence, a tremendous amount of works on enhancing program
representation learning and
comprehension emerges recently~\cite{AllamanisBDS18,abs-2002-05442}. 

Some recent works also study modeling ASTs and scale training to large programs. 
Allamanis et al.~\cite{AllamanisBK18} represent programs as graphs. Gated Graph Neural
Network~\cite{LiTBZ15} is adopted to learn the representation of programs for variable misuse and variable naming prediction. 
Alon et al.~\cite{Alon2019Code2vec} decompose an AST as a collection of paths and generate the representations of code by learning how to aggregate paths.
Wang and Li~\cite{WangL21} tailor GNN to modeling AST for code completion task.
Zhang et al.~\cite{ZhangWZ0WL19} split an AST into statements
and use a preorder traversal to transform the AST to a sequence of statements.
Then, the statement sequence is fed into RvNN to construct the statement representations which
are finally passed to GRU for code classification and clone detection. 
Although their idea resembles our work, they decompose ASTs based on individual statements. This way, the information beyond the boundary of statements is hard to capture.

Inspired by the success of the pre-trained model BERT~\cite{DevlinCLT19}, pre-training has also been applied in program representation learning.  
CodeBERT~\cite{FengGTDFGS0LJZ20} is a bimodal pre-trained model on programming
language and natural language, and it is designed to benefit down-stream tasks. 
However, CodeBERT produces poorer results compared to methods which make use of AST paths~\cite{FengGTDFGS0LJZ20} since it only takes original code as the input.

\section{Conclusion}\label{sec:conclusion}
In this paper, we have presented a novel method \ours to improve code
summarization. 
To alleviate the problem of difficult and slow training on ASTs, 
\ours uses block-wise code splitting and localizes the hierarchical syntax
structure of AST generated for each code split. \ours further 
adopts a pre-training strategy to capture local non-linear syntax
encoding of split ASTs, which helps its Transformer-based summarization model generate high-quality code summaries.
Experiments on public benchmarks have demonstrated that \ours
significantly outperform state-of-the-art models. In the
future, we plan to exploit more sophisticated program slicing techniques and
combine code slicing based on variables with variable-aware neural
summarization models to further improve \ours. 

\clearpage

\section*{Acknowledgment}
Chen Lin is supported by the National Natural Science Foundation of China (No. 61972328), and Joint Innovation Research Program of Fujian Province of China (No. 2020R0130).
Hui Li is supported by the National Natural Science Foundation of China (No. 62002303, 62036004), and the Natural Science Foundation of Fujian Province of China (No. 2020J05001).
Rongxin Wu is supported by the National Natural Science Foundation of China (No. 61902329).

\bibliographystyle{IEEEtran}
\bibliography{main}

\end{document}